\documentclass[3p,times]{elsarticle}

\usepackage{ecrc}

\usepackage{graphicx,epstopdf}
\usepackage{amsmath,amssymb}

\volume{00}

\firstpage{1}

\journalname{Nuclear Physics A}

\runauth{}

\jid{nupha}

\usepackage{amssymb}

\usepackage[figuresright]{rotating}

\begin{document}

\begin{frontmatter}



\dochead{}

\title{Jet-Tagged Back-Scattering Photons For Quark Gluon Plasma Tomography}


\author[a]{Rainer J.\  Fries}
\ead{rjfries@comp.tamu.edu}
\author[a,b]{S.\ De}
\author[b]{D.\ K.\ Srivastava}

\address[a]{Cyclotron Institute and Department of Physics and Astronomy, Texas
  A\&M University, College Station, TX 77845, USA}
\address[b]{Variable Energy Cyclotron Centre, 1/AF, Bidhan Nagar, Kolkata
  - 700064, India}

\begin{abstract}
Several sources of direct photons are known to contribute to the total photon
yield in high energy nuclear collisions. All of these photons carry characteristic
and important  information on the initial nuclei or the hot and dense fireball
created in the collision. We investigate  the possibility to separate photons
from back-scattering of high momentum quarks off quark gluon plasma from
other sources. Their unique kinematics can be utilized through high energy jet 
triggers on the away-side. We discuss the basic idea and estimate the
feasibility of  such a measurement at RHIC and LHC.
\end{abstract}


\end{frontmatter}



\section{Introduction}
\label{sec:intro}

Photons and dileptons hold great promise as penetrating probes of hot nuclear
matter. Their mean free path is short compared to the typical size of the
fireball created in high energy nuclear collisions. Hence, once they are
created their rescattering can be neglected and they encode valuable
information on their creation process. 
Different sources of direct photons have been discussed in the literature over
the years. (i) {\it Hard photons from initial scatterings of partons
in the nuclear wave functions}. They can be used to extract information about
the modification of nucleon wave functions in large nuclei  \cite{Owens:1986mp}.
(ii) {\it Photons fragmenting off jets created in initial hard collisions}. Together
with the previous source these photons are already present in elementary $p+p$
collisions. Fragmentation photons, besides being sensitive to the
same wave function effects as hard initial photons can be suppressed through 
energy loss of the partons before the photon is radiated off
\cite{Owens:1986mp}. Hard initial and 
fragmentation photons can in principle be distinguished experimentally through
isolation cuts which check for hadronic activity in a cone around the photon.
(iii) {\it Pre-equilibrium photons}. This poorly understood photon source is
active during the first fm/$c$ of a nuclear collision when partons are
reinteracting but are not thermalized \cite{Bass:2002pm}.
(iv) {\it Thermal QGP radiation}. Thermal emission from an equilibrated quark
gluon plasma (QGP) is one of the signature measurements in heavy ion physics
\cite{Kapusta:1991qp,Baier:1991em,Aurenche:1998nw,Arnold:2001ms}, as it could serve 
as proof of deconfinement and as a thermometer for the matter created 
\cite{Adare:2008ab}. Thermal QGP radiation might however not be easily
distinguishable from other thermal emission. (v) {\it Thermal radiation from 
hot hadronic gas}. Hadronic emission is an important contribution to the
overall yield at low photon momenta \cite{Kapusta:1991qp}. Albeit at lower 
temperature than the QGP photons the rate does not vary much across the pseudo-critical 
temperature $T_c$ for deconfinement.
(vi) {\it Photons from the hadronization processes}.  Photons from hadronization 
of QGP --- while undoubtedly present \cite{ChenFries:2012} --- have largely 
been ignored in the literature so far due to the intrinsically
non-perturbative nature of the process. (vii) {\it Photons from jets
  interacting with the medium}. It has been argued that jet energy loss in
QGP and in hot hadronic matter should be accompanied by photon back-scattering
\cite{Fries:2002kt,Fries:2005zh}  and bremsstrahlung \cite{Zakharov:2004bi}. Back-scattering photons are sensitive to
the mean free path of fast partons in QGP and encode information about parton 
energy loss that is complementary to hadronic measurements.
In these proceedings we propose a novel way to measure back-scattering
photons from fast partons in QGP.

\section{Back-Scattering Photons}
\label{sec:bs}

It is well known in electrodynamics that elastic photon-electron scattering
exhibits a sharp peak in backward direction. This Compton back-scattering
phenomenon has been used extensively to create collimated beams of high
energy photons by scattering an intense laser pulse ($E_\gamma \sim 1$ eV)
off a beam of high energy electrons ($E_{e^{-}} \sim 1$ -- 100 GeV)
 \cite{Milburn:1962jv,Arutyuninan:1963aa}. In quantum chromodynamics
the mixed process $q + g \to q +\gamma$ involves the same type of diagrams
at leading order and exhibits the same Compton back-scattering peak. A similar
peak can be observed for the annihilation process $q+\bar q \to \gamma+ g$.
Hence we expect a significant yield of high energy photons from quarks 
($E_q \sim 1$ -- 100 GeV) scattering off thermal gluons ($E_g \sim 200$ MeV)
or thermal antiquarks.

In \cite{Fries:2002kt} some of us have estimated the rate of high energy Compton
back-scattering and annihilation photons from jets interacting with the medium
to be
\begin{equation}
\label{eq:1}
  E_\gamma \frac{dN}{d^4x d^3p_\gamma} = \frac{\alpha\alpha_s}{4\pi^2}
  \sum_{f=1}^{N_f} \left(\frac{e_f}{e}\right)^2 
  \left[ f_q(\mathbf{p}_\gamma,x) + f_{\bar q}(\mathbf{p}_\gamma,x) \right]
  T^2 \left[ \ln \frac{3E_\gamma }{\alpha_s\pi T} + C\right]
\end{equation}
where $C=-1.916$. $\alpha$ and $\alpha_s$ are the electromagnetic and strong 
coupling constant, $T$ is the local temperature at $x$, $f_q$ is the phase
space distribution of fast quarks interacting with the medium and $e_f$ is the 
electric charge of a quark with the index $f$ running over all active quark flavors. 
This rate has subsequently also been calculated for virtual photons 
\cite{Srivastava:2002ic,Turbide:2006mc}.

We note a few interesting facts about the back-scattering rate. First, it is
parametrically proportional to $T^2 \ln 1/T$ and hence very sensitive to the
medium temperature. Secondly, it is proportional to the quark distribution
$f_q$ as a function of momentum, position and time. Hence back-scattering
photons have a power-law like spectrum that can make a sizable 
contribution even at intermediate and large photon transverse momentum $P_T$.
Nevertheless the normalization of this spectrum is still sensitive to the
temperature of the medium. This leads to the interesting prospect of finding
thermal signatures at large momenta. Furthermore, because the quark
distributions evolve with time the back-scattering spectra are sensitive to
quark energy loss.

In Refs.\ \cite{Fries:2002kt,Fries:2005zh,Turbide:2007mi} it has been noted
that jet-medium photons can make a sizable contribution to the total photon
yield both at the Relativistic Heavy Ion Collider (RHIC) and the Large
Hadron Collider (LHC). They shine particularly bright compared to other
sources in the intermediate $P_T$ range around 4 GeV/$c$. This comes from
the power-law behavior which lifts them above purely thermal sources at those
momenta, while on the other hand there is an additional $\sim 1/P_T$ factor 
\cite{Fries:2002kt} that makes the power-law slightly softer than the initial hard
photon spectrum which dominates at very large $P_T$.
However, it is very difficult to experimentally confirm the presence
of jet-medium photons, let alone produce a quantitative measurement, by
analyzing singe inclusive photon spectra alone.
Subsequently it was proposed by some of us to use the azimuthal asymmetry
$v_2$ around the beam axis as a signature \cite{Turbide:2005bz}. This is based
on the observation that jet-medium photons should have negative $v_2$ unlike
any of the other photon sources whose $v_2$ is positive or vanishing 
\cite{Chatterjee:2005de}. However, the small absolute size of direct photon
$v_2$ at intermediate and large momenta has made this impossible with
currently available data \cite{Turbide:2007mi}.

Here we propose a novel way to experimentally confirm the existence of
jet-medium photons and maybe extract quantitative information from them.
The basic idea starts with a classification of hadron-photon or jet-photon 
correlations inherent in the sources discussed above. It is clear that only
initial hard photons, fragmentation photons and jet-medium photons have
a significant back-to-back correlation with hadrons or jets. We will focus on
jets in the following which lead to cleaner signals and are now routinely
measured at LHC and RHIC. Thus by measuring photons on the opposite side of
a trigger jet one can eliminate most competing photon sources right away.
Next we note that fragmentation photons are concentrated at low $z$ ($\lesssim 0.3$)
where $z=E_\gamma/E_{\mathrm{parent\,\, jet}}$
\cite{Owens:1986mp,Bourhis:2000gs} .  
Hence background from this source can be suppressed by choosing a kinematics 
which features large $E_\gamma$, of the order of the trigger jet energy.

At leading order (LO) kinematics the transverse momenta of the trigger jet and an
associated hard initial photon balance perfectly,$P_{T,\gamma} =
P_{T,{\mathrm{trigger}}}$. Without energy loss of the partons before radiation 
the back-scattering kinematics would ensure the same momentum correlation with
the trigger for back-scattering photons, potentially burying them under the hard direct
photon peak. However, if the leading parton of the associated jet suffers from energy
loss, the back-scattering photon $P_T$ will be shifted away from the
trigger $P_T$, typically a few GeV below $P_{T,{\mathrm{trigger}}}$. This shift
by itself, if measured, carries valuable information about parton energy
loss. At next-to-leading order (NLO) kinematics the back-to-back correlation
and the strong momentum correlations get washed out but residual signals remain.

\section{Results}
\label{sec:results}

In this section we present the results of some preliminary studies of the
signal (back-scattering photons) and the background (initial hard and
fragmentation photons).
We use the JETPHOX code \cite{Catani:2002ny,Aurenche:2006vj} for jet-photon
and jet-hadron yields at LO and NLO to calculate the underlying hard process
and the background from initial hard and fragmentation photons. The trigger jets were 
fixed in trigger windows around midrapidity, then the photon spectra were 
calculated at midrapidity in sectors of $30^\circ$ width around the away-side.
We use the PPM code \cite{Rodriguez:2010di,Fries:2010jd} to propagate leading 
jet partons of the away-side jet inside a longitudinally boost-invariant
fireball with a $L^2$-dependent energy loss (sLPM in
\cite{Rodriguez:2010di}) and to convert them to photons according to Eq.\ (\ref{eq:1}).
The energy loss parameter in PPM is fixed to describe measured single
inclusive hadron spectra.

\begin{figure}[t]  
\begin{center}
  \includegraphics[width=8cm]{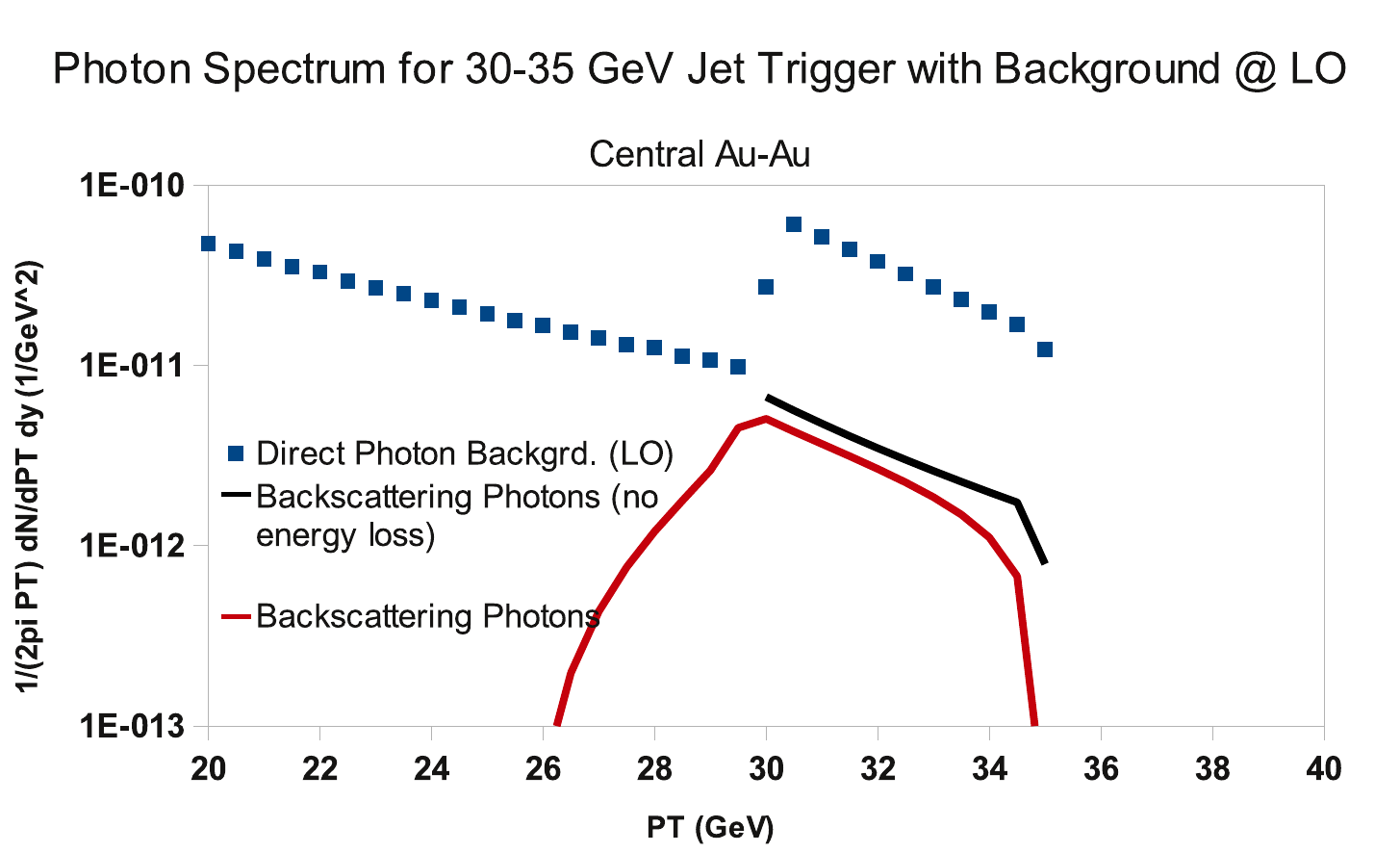}
  \includegraphics[width=8cm]{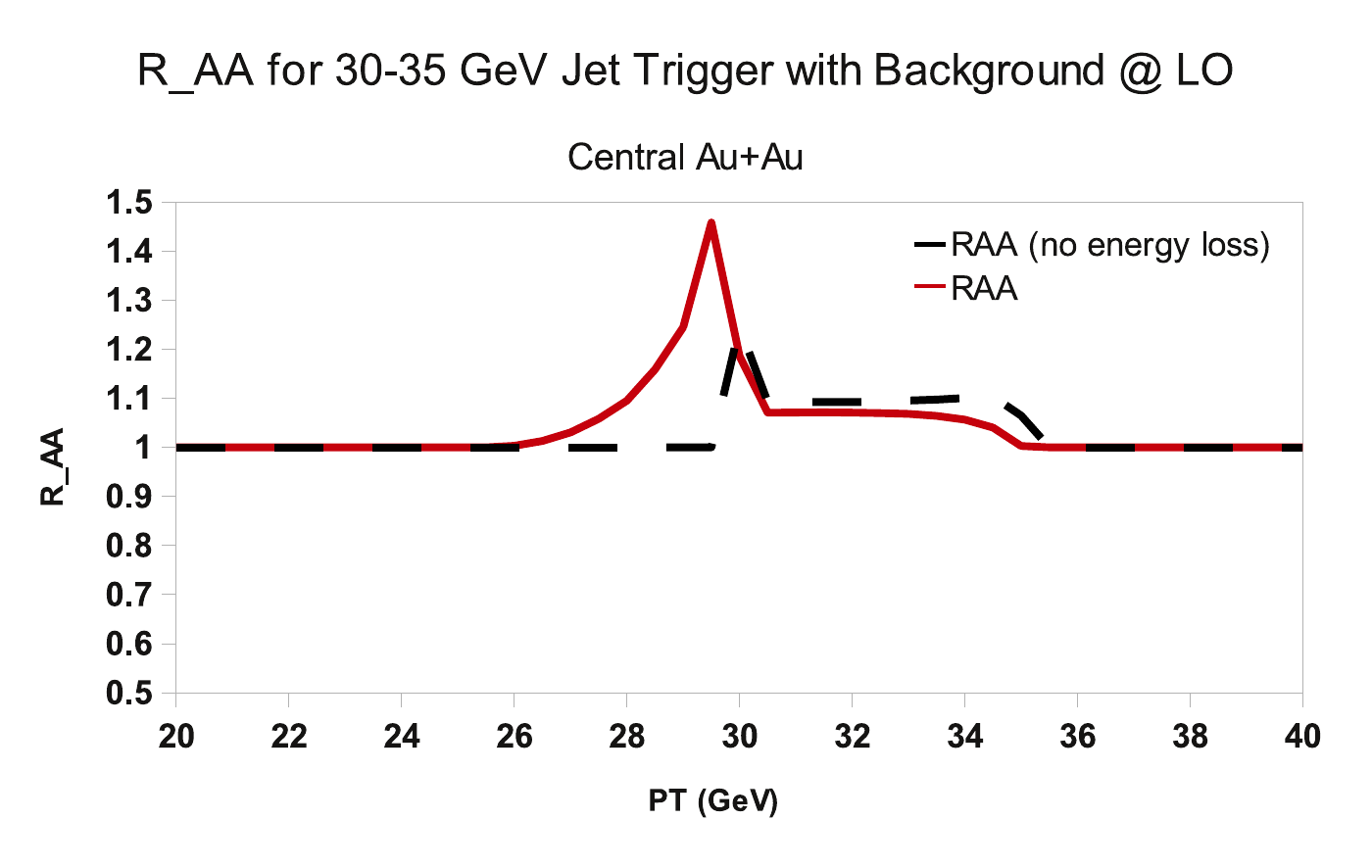}
  \caption{\label{fig:1}
    Left panel: photon spectra associated with trigger jets from 30-35
    GeV in central Au+Au collisions at RHIC at $\sqrt{s_{NN}}=200 $ GeV. The
    yield of back-scattering photons with (red solid line) and without (black dashed
    line) energy loss is compared to the background
    of initial hard photons and fragmentation photons (blue squares). Right
    panel: the nuclear modification factor $R_{AA}$ with (red solid line) and
    without (black dashed line) taking energy loss of leading partons into account.}
\end{center}
\end{figure}

\begin{figure}[t] 
\begin{center}
  \includegraphics[width=8cm]{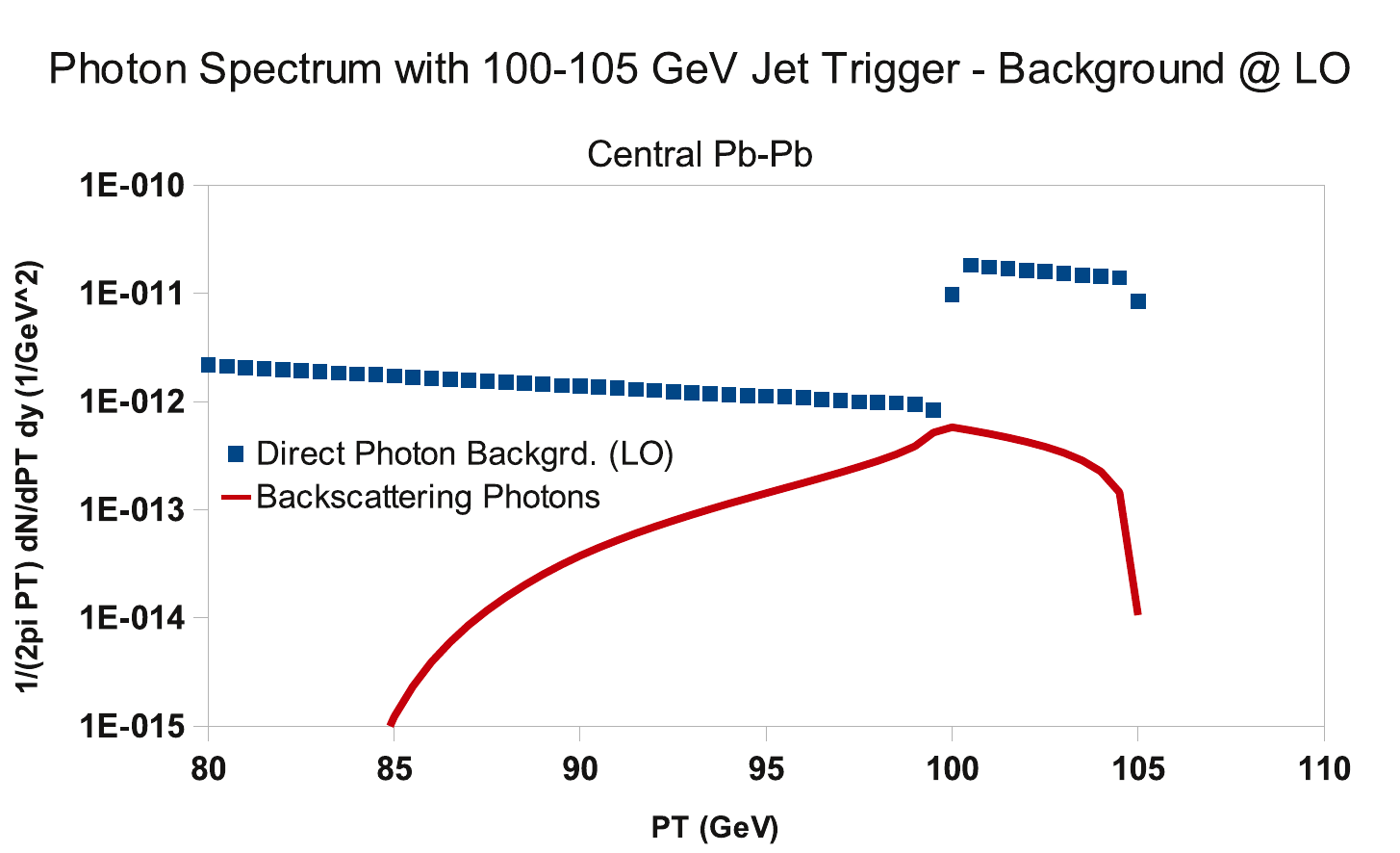}
  \includegraphics[width=8cm]{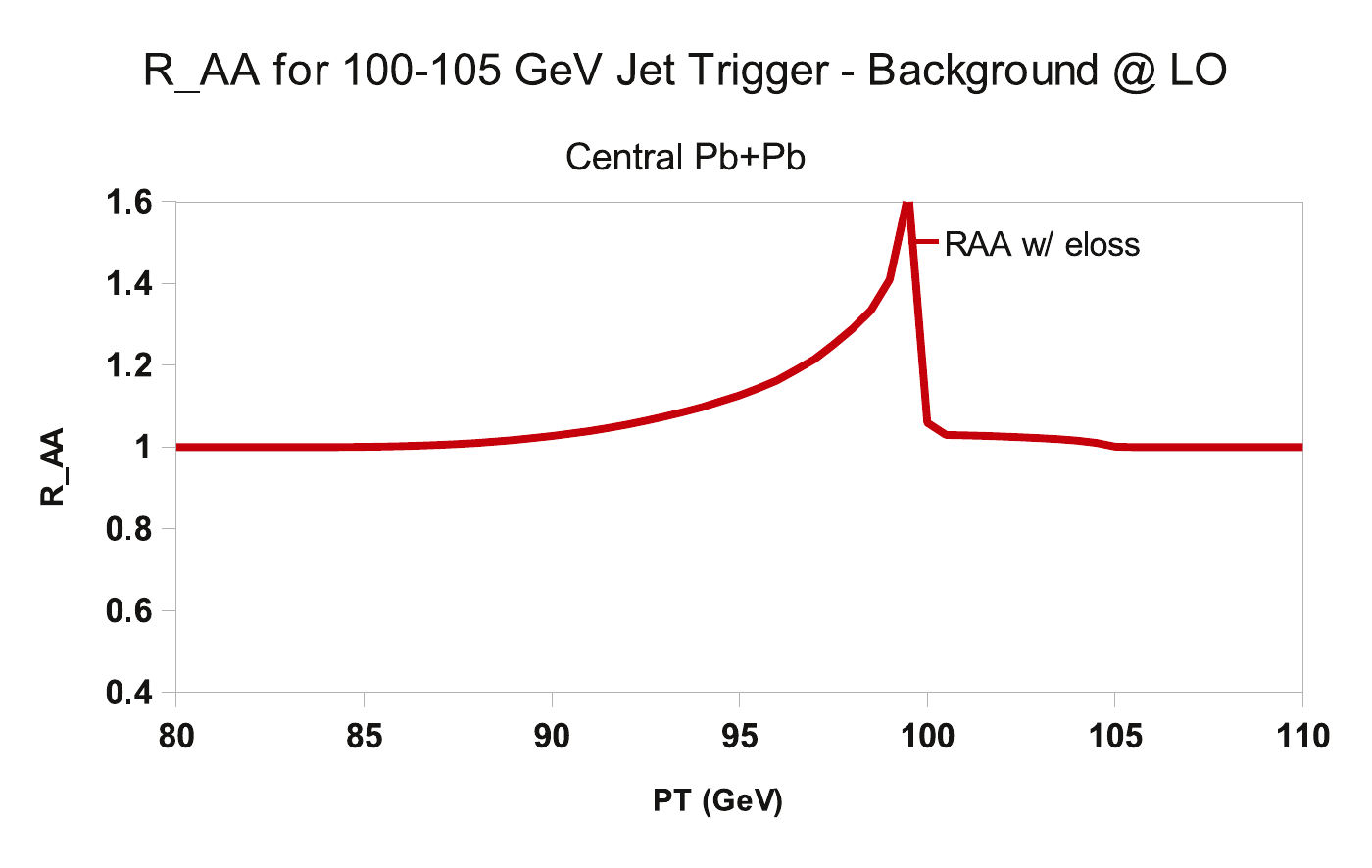}
  \caption{\label{fig:2}
    The same as Fig.\ \ref{fig:1} for 100-105 GeV trigger jets in central Pb+Pb
  collisions at the LHC at $\sqrt{s_{NN}}=2.76 $ TeV. Only the (realistic)
  case of back-scattering photons with energy loss prior to their emission is shown.}
\end{center}
\end{figure}

Fig.\ \ref{fig:1} shows the results for central Au+Au collisions at RHIC
($\sqrt{s_{NN}}=200 $ GeV) for a jet trigger window of $E_T = 30$-$35$ GeV for
LO kinematics. Direct hard photons exhibit a distinctive peak inside the
trigger window while fragmentation photons slowly rise from the trigger window
towards lower $z$. Back-scattering photons without energy loss prior to the
Compton or annihilation process are strongly correlated with the trigger
window as well. However, if energy loss is switched on they show a distinct 
shoulder toward smaller momenta due to energy loss of partons before the 
Compton or annihilation process. We also
show the nuclear modification factor $R_{AA}$ which is approximated by the 
ratio of back-scattering photons (signal) over sum of back-scattering (signal)
and hard initial and fragmentation (background) photons. A signal can be seen
across the trigger window. A characteristic peak develops once energy loss
is switched on. Fig.\ \ref{fig:2} shows the same for central Pb+Pb
collisions at the LHC ($\sqrt{s_{NN}}=2.76 $ TeV) for a jet trigger window of 
$E_T = 100$-$105$ GeV.
Initial results for backgrounds at NLO accuracy show  a decorrelation
between trigger and and hard initial photon momentum as expected. This
might decrease the signal-to-background by up to a factor two.

\section{Summary}
\label{sec:summary}

Separating and quantitatively using different direct photon sources in high
energy nuclear collisions is challenging. We advocate the use of jet triggers
to separate back-scattering photons from a background of fragmentation and
hard initial photons. We predict that back-scattering photons produce a 
characteristic enhancement in nuclear modification factors for direct photons 
on the away side of the trigger, just below the trigger jet energies. This 
technique completely subtracts all direct photon sources without an 
underlying hard process. It further suppresses fragmentation photons 
which amass at low $z$, and hard direct photons which do not suffer 
from energy loss. The height of the back-scattering peak reflects a 
quadratic temperature dependence and the shift of the peak down 
from the trigger window is sensitive to parton energy loss.
In the future one should study the effects of trigger jet energy loss which
has been neglected so far, as well as add a next-to-leading order calculation of
the hard process underlying the back-scattering signal.

This work was supported by NSF CAREER Award PHY-0847538 and by the JET
Collaboration and DOE grant DE-FG02-10ER41682.








\end{document}